\documentclass[aps,prb,twocolumn,superscriptaddress,showpacs]{revtex4}
\usepackage{graphicx}

\begin{document}


\title{Temperature Dependent Magnetic Anisotropy in (Ga,Ma)As Layers}


\author{M. Sawicki}
 \email{mikes@ifpan.edu.pl}
\affiliation{Institute of Physics, Polish Academy of Sciences, al.
Lotnik\'ow 32/46, 02-668 Warszawa, Poland}
\author{F. Matsukura}
\affiliation{Institute of Physics, Polish Academy of Sciences, al.
Lotnik\'ow 32/46, 02-668 Warszawa, Poland} \affiliation{Laboratory
for Electronic Intelligent Systems, Research Institute of
Electrical Communication,\\ Tohoku University, Katahira 2-1-1,
Sendai 980-8577, Japan}
\author{A. Idziaszek}
\affiliation{Institute of Physics, Polish Academy of Sciences, al.
Lotnik\'ow 32/46, 02-668 Warszawa, Poland}
\author{T. Dietl}
\affiliation{Institute of Physics, Polish Academy of Sciences, al.
Lotnik\'ow 32/46, 02-668 Warszawa, Poland}
\author{\\G.M. Schott}
\affiliation{Physikalisches Institut, Universit\"at W\"urzburg, Am
Hubland, D-97074 W\"urzburg, Germany}
\author{C. Ruester}
\affiliation{Physikalisches Institut, Universit\"at W\"urzburg, Am
Hubland, D-97074 W\"urzburg, Germany}
\author{G. Karczewski}
\affiliation{Institute of Physics, Polish Academy of Sciences, al.
Lotnik\'ow 32/46, 02-668 Warszawa,
Poland}\affiliation{Physikalisches Institut, Universit\"at
W\"urzburg, Am Hubland, D-97074 W\"urzburg, Germany}
\author{G. Schmidt}
\affiliation{Physikalisches Institut, Universit\"at W\"urzburg, Am
Hubland, D-97074 W\"urzburg, Germany}
\author{L.W. Molenkamp}
\affiliation{Physikalisches Institut, Universit\"at W\"urzburg, Am
Hubland, D-97074 W\"urzburg, Germany}

\date{\today}

\begin{abstract}
It is demonstrated by SQUID magnetization measurements that
(Ga,Mn)As films can exhibit rich characteristics of magnetic
anisotropy depending not only to the epitaxial strain but being
strongly influenced by the hole and Mn concentration, and
temperature. This behavior reflects the spin anisotropy of the
valence subbands and corroborates predictions of the mean field
Zener model of the carrier mediated ferromagnetism in III-V
diluted magnetic semiconductors with Mn. At the same time the
existence of in-plane uniaxial anisotropy with [110] the easy axis
is evidenced. This is related to the top/bottom symmetry breaking,
resulting in the lowering of point symmetry of (Ga,Mn)As to the
C$_{2v}$ symmetry group. The latter mechanism coexists with the
hole-induced cubic anisotropy, but takes over close to
$T_{\mbox{\tiny C}}$.
\end{abstract}


\maketitle

The discovery of carrier-mediated ferromagnetism at temperatures
in excess of 100~K in (III,Mn)V dilute magnetic semiconductors
(DMS) grown by molecular beam epitaxy (MBE) has made it possible
to combine complementary properties of semiconductor quantum
structures and ferromagnetic systems in single devices, paving the
way for the development of functional semiconductor
spintronics.\cite{Spintronics} Therefore, the understanding of
magnetic anisotropy in these systems and the demonstration of
methods for its control is timely and important. It has been
known, since the pioneering works of Munekata {\it et
al.}\cite{Mune93} and Ohno {\it et al.},\cite{Ohno96Shen97} that
ferromagnetic (In,Mn)As and (Ga,Mn)As films are characterized by a
substantial magnetic anisotropy. Remarkably, due to magnetic
dilution, the ordinary shape anisotropy plays here only a marginal
role and, accordingly, explains neither direction nor large
magnitude of the observed anisotropy field $H_{un}$.

It has been found by anomalous Hall effect
studies,\cite{Ohno96Shen97} that the direction of the easy axis is
mainly controlled by epitaxial strain in these systems. Generally,
for layers under tensile biaxial strain [like (Ga,Mn)As on a
(In,Ga)As buffer] perpendicular-to-plane magnetic easy axis has
been observed (perpendicular magnetic anisotropy, PMA). In
contrast, layers under compressive biaxial strain [as canonical
(Ga,Mn)As on a GaAs substrate] have been found to develop in-plane
magnetic easy axis (in-plane magnetic anisotropy, IMA). At first
glance, this sensitivity to strain appears surprising, as the Mn
ions are in the orbital singlet state $^6A_1$. For such a state,
the strain-induced single ion anisotropy is expected to be rather
small and, indeed, electron paramagnetic resonance (EPR) studies
of Mn in GaAs yielded relevant spin Hamiltonian parameters by two
orders of magnitude too small to explain the observed values of
$H_{un}$.\cite{Fedo02}

In the system in question, however, the ferromagnetic spin-spin
exchange interaction is mediated by the band holes, whose
Kohn-Luttinger amplitudes are primarily built up of anion p
orbitals in tetrahedrally coordinated semiconductors. Furthermore,
in semiconductors, in contrast to metals, the Fermi energy is
usually smaller than the atomic spin-orbit energy. Hence, as noted
by some of the present authors and co-workers,\cite{Diet97,Diet00}
the strain or confinement-induced anisotropy of the valence band
can result in a sizable anisotropy of spin properties. Indeed, the
quantitative calculation within the mean-field Zener
model,\cite{Diet00,Diet01a} in which the valence band is
represented by the 6$\times$6 Luttinger hamiltonian, explains the
experimental values of $H_{un}$ in (Ga,Mn)As with the accuracy
better than a factor of two. Moreover, by combining theories of
magnetic anisotropy\cite{Diet00,Diet01a,Abol01} and of magnetic
stiffness,\cite{Koni01} it has been possible to describe the width
of stripe domains in (Ga,Mn)As PMA films.\cite{Diet01b} However,
the theories in question\cite{Diet01a,Abol01} contain a number of
predictions that call for a detail experimental verification.

In this communication we present magnetic anisotropy studies
carried out by direct magnetization measurements in a dedicated
SQUID magnetometer.  Our results demonstrate that the
temperature-induced reorientation of the easy axis from [001] to
[100], and then to [110] or equivalent directions occurs in films
of (001) (Ga,Mn)As on GaAs with appropriately low values of $p$.
We argue that these findings corroborate the above mentioned
theoretical expectations. However, the data collected near
$T_{\mbox{\tiny C}}$ reveal a non-equivalence of the $[110]$ and
$[\bar{1}10]$ crystal directions, the latter corresponding to the
hard axis. This points to the existence of a mechanism lowering
the point symmetry from D$_{2d}$ to C$_{2v}$. We find that this
particular anisotropy develops independently of $p$. Such a
symmetry breaking has already been inferred from transport data
\cite{Kats98,Gall02} and Kerr rotation.\cite{Hrab02} Finally, we
find that in another class of samples the hard axis remains
oriented along the $[\bar{1}10]$ direction but the three other
main directions ([100], [110] and [010]) become equally easy. We
discuss reasons and consequences of these unusual characteristics
of magnetic anisotropy in (Ga,Mn)As.

Two series of (Ga,Mn)As films with the thickness of 350 and 500~nm
were coherently deposited by MBE onto (100) GaAs substrates under
As$_4$/Ga beam equivalent pressure (BEP) ratio of 5 and 25, and
growth temperature $T_{sub}$~=~220 or 270$^{\mbox{\small{o}}}$C.
The Mn composition $x$ varies from 0.007 to 0.067, as established
by x-ray diffraction measurements.\cite{Scho01} High resolution
x-ray diffraction shows high crystal quality of (Ga,Mn)As with the
rocking curve widths comparable to those of the GaAs substrate and
pronounced finite-thickness fringes, indicating flat interfaces
and surfaces. The layers are pseudomorphic with respect to the
GaAs substrate, as verified by reciprocal space maps around the
asymmetric [115] reflex.

Magnetization measurements are carried out in a home made
superconducting quantum interference device (SQUID) magnetometer
down to 5~K and in the magnetic fields up to 3~kOe. By design, the
signal detection axis of the magnetometer (denoted in Fig.~1 as
a$_m$), is aligned with the direction of the magnetic field
\emph{H} (vertically). As a result the system is sensitive only to
the vertical component of the magnetization vector. A special care
is put forward to screen the sample from external magnetic fields
and to keep the parasite residual fields generated by the magnet
at the lowest possible level. The Meissner effect of a pure lead
sample proves that even after excursion to the field of 1~kOe, the
residual field remains below 60 mOe. Such a low value is essential
for successful magnetic remanence studies or for measurements at
temperatures close to $T_{\mbox{\tiny C}}$, where the coercivity
of (Ga,Mn)As drops into a sub-Oersted range. For studies of
in-plane anisotropy, (Ga,Mn)As layers are shaped by chemical
etching into circles of about 5~mm in diameter.

\begin{figure}   
\includegraphics[width=7cm]{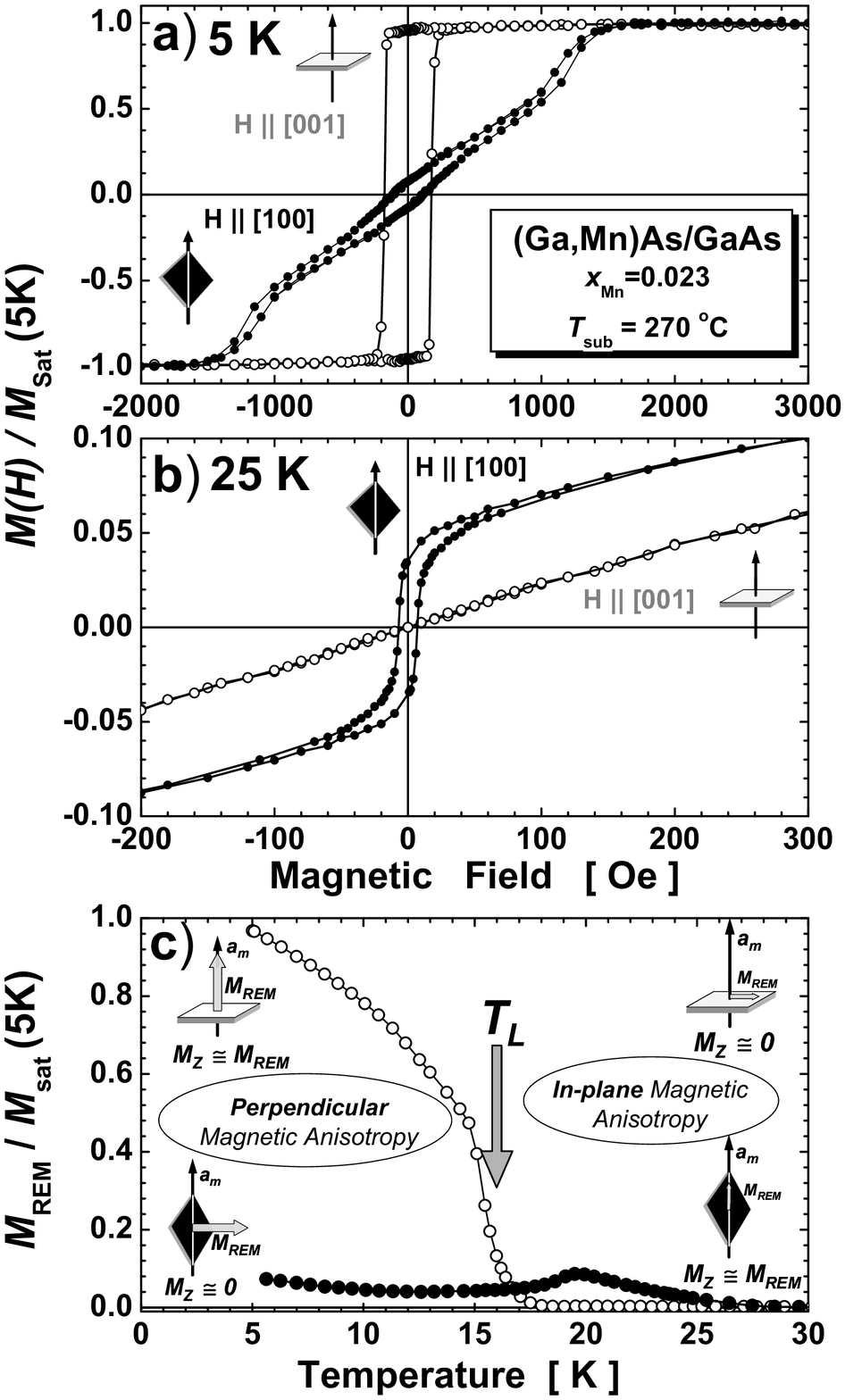}
\caption{\label{fig1}Magnetic field (a,b) and temperature (c)
dependent magnetization measured for parallel (full points) and
perpendicular (open points) orientations with respect to the
magnetizing field. Note the flip of the easy axis at $T_{\mbox
{\tiny L}}$.}
\end{figure}

Figures 1(a) and 1(b) present typical hysteresis loops for a
Ga$_{0.977}$Mn$_{0.023}$As sample, which exhibits the temperature
driven change from PMA to IMA. At low temperatures, Fig.~1(a), a
perfect square hysteresis is obtained when the magnetic field is
perpendicular to the film surface, while an elongated loop is seen
when the in-plane orientation is probed. This clearly evidences
that the easy axis is perpendicular. Remarkably, a reverse behavior
is observed at higher temperatures, Fig.~1(b). These findings
demonstrate that the easy axis flips from the perpendicular to
in-plane orientation when the temperature increased.

In order to trace directly the reorientation of spontaneous
magnetization, we have examined the temperature dependence of
remanent magnetization $M_{\mbox{\tiny {REM}}}$ along selected
crystallographic directions according to the following procedure.
After being oriented on the holder, the sample is cooled down
through $T_{\mbox{\tiny C}}$\ in the field \emph{H}$_{\mbox{\tiny
FC}}=1$~kOe, which is at least a few times larger than the
coercive field $H_c$. Then, the field is removed at 5~K, and the
measurement of the $M_{\mbox{\tiny {REM}}}$ component along the
direction of \emph{H}$_{\mbox{\tiny{FC}}}$ commences on increasing
temperature in the residual field \emph{H}$_r\approx 60$~mOe.

As shown in Fig.~1(c), the $M_{\mbox{\tiny{REM}}}$ signal at low
temperatures is tenfold larger for the perpendicular orientation,
\emph{H}$_{\mbox{\tiny FC}}\parallel [001]$, comparing to the case
of parallel arrangement, \emph{H}$_{\mbox{\tiny {FC}}}
\parallel [100]$. This reconfirms the appearance of PMA for large
values of spontaneous magnetization $M_s$ in this film despite the
presence of a sizable compressive strain. The PMA holds until a
certain temperature $T_{\mbox {\tiny L}}$ ($T_{\mbox{\tiny L}}
\simeq16$~K in this case), at which a sudden drop of the signal is
detected in the perpendicular orientation. Nevertheless, the
$M_{\mbox{\tiny {REM}}}$ signal for the parallel case remains
small. This seems to contradict our scenario, according to which
the easy axis is in-plane for $T>T_{\mbox {\tiny L}}$. We note,
however, that the spontaneous flip of magnetization to the
in-plane configuration results in a {\em demagnetized} state.
Actually, a visible maximum of the parallel $M_{\mbox{\tiny
{REM}}}$ between $T_{\mbox{\tiny L}}$ and $T_{\mbox{\tiny C}}$
reflects the presence of the field \emph{H}$_r$, which magnetizes
the film once the in-plane anisotropy and coersive fields become
sufficiently small. Indeed, a temperature cycling of about $\Delta
T = 2$~K just below this maximum increases the magnitude of the
in-plane magnetization by a factor of three, as shown in Fig.~2.

\begin{figure}   
\includegraphics[width=7cm]{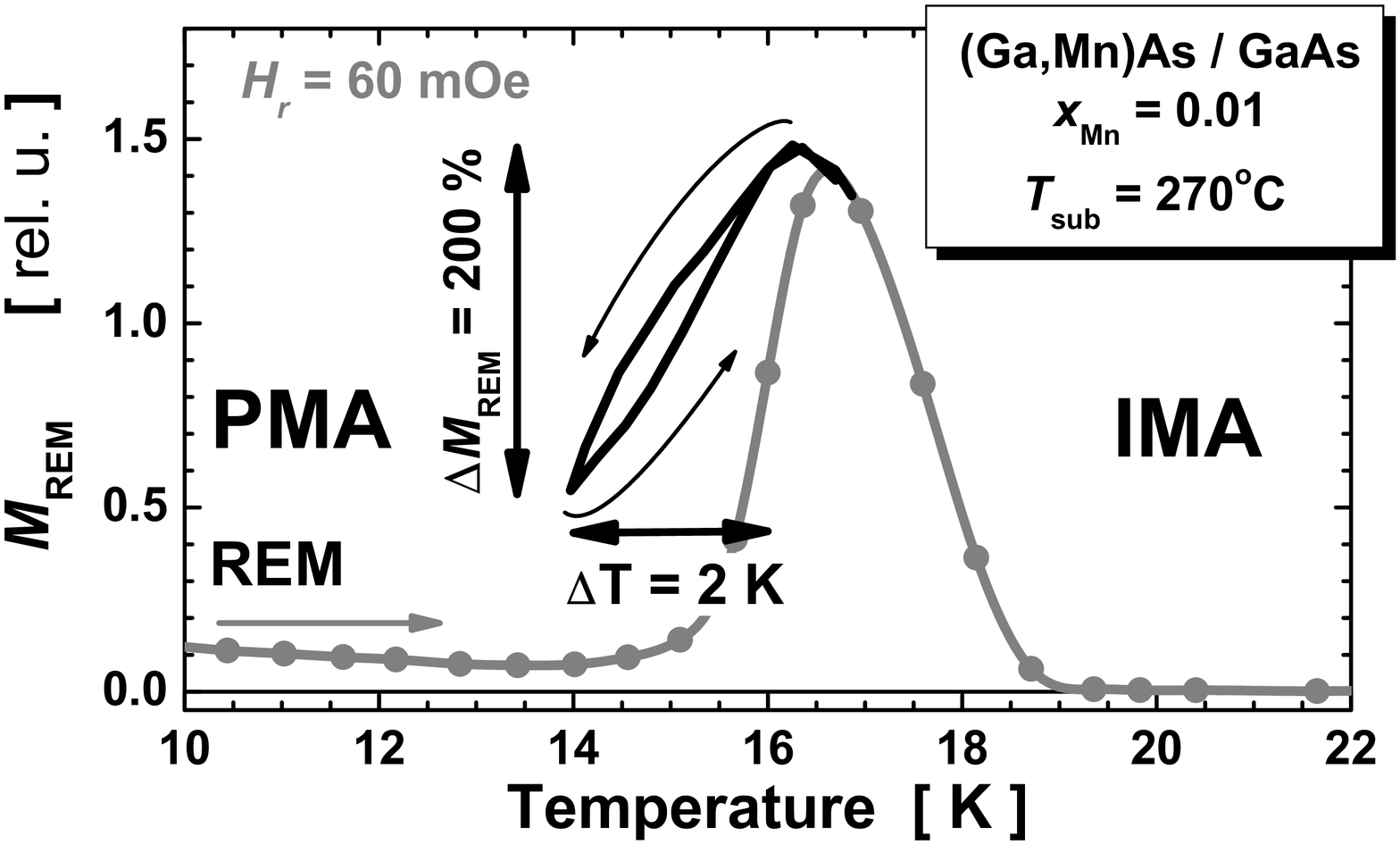}
\caption{\label{fig2}Temperature cycling of the remanent in-plane
magnetization in Ga$_{0.99}$Mn$_{0.01}$As at the border of
perpendicular to in-plane anisotropy conversion.}
\end{figure}

We have found the temperature-induced cross over from PMA to IMA
described above in five other samples of (Ga,Mn)As.\cite{Sawi02}
Recently, this effect has been detected in (Al,Ga,Mn)As under
similar conditions.\cite{Taka02}  According to low temperature
resistance measurements, such samples exhibit an insulating
character, which points to rather small hole concentrations $p$, a
conclusion consistent with relatively low magnitudes of the Curie
temperature in these films, $T_{\mbox{\tiny C}} < 30$~K.
Remarkably, the opposite behavior occurs in (In,Mn)As under a
tensile strain, where the easy axis flips from the in-plane to the
out of plane on warming.\cite{Endo01} We assign the temperature
dependent anisotropy to the repopulation of the valence band
subbands, which occurs when the Mn magnetization changes.  For a
qualitative analysis we refer to numerical predictions depicted in
Fig.~10 of Ref.~\onlinecite{Diet01a}. It is expected that for
compressive (tensile) strain the easy axis changes direction from
the in-plane to perpendicular orientation on lowering (raising)
$p$, the critical value being of the order of $p_c
=5\times10^{19}$~cm$^{-3}$ for $M_s$ corresponding to the
saturation value of magnetization for $x = 0.02$. We conclude that
the theoretically expected value of $p_c$ is consistent with our
results for (Ga,Mn)As/GaAs, which demonstrate the existence of PMA
only in samples with sufficiently small values of $p$.
Furthermore, at given $p$, a cross-over from the PMA to IMA is
expected on decreasing $M_s$, that is on increasing $T$, as
observed. This temperature-driven reorientation of the easy axis
reflects the equipartition  of the valence subband population at
$T$ approaching $T_{\mbox{\tiny C}}$ and, therefore, reconfirms
the crucial role of the valence band holes in the ferromagnetism
of (III,Mn)V systems. Unfortunately, however, a more quantitative
comparison is hampered by the well-known difficulties in the
accurate determination of $p$ in films of (Ga,Mn)As, particularly
near the metal-insulator transition.

We now turn to (Ga,Mn)As films which, owing to sufficiently large
$p$, exhibit IMA in the whole temperature range above 5~K. In this
case, according to the theoretical predictions presented in Fig.~9
of Ref.~\onlinecite{Diet01a} and in Fig.~6 of
Ref.~\onlinecite{Abol01}, the easy axis is expected to switch
between $[100]$ and $[110]$ or equivalent in-plane directions as a
function of $p$. Alternatively, at given $p$, the easy direction
may change when $M_s$ is varied, say, by $x$ or $T$. In fact,
there exists a class of samples, in which we observe the
temperature-induced change of the IMA. Data collected in Fig.~3,
by presenting $M_{\mbox{\tiny {REM}}}$ measured along the four
principal in-plane crystallographic directions $[100]$, $[110]$,
$[010]$, and $[\bar{1}10]$, exemplify this case. We note that at
low temperatures the highest values of $M_{\mbox{\tiny {REM}}}$
are collected along $\langle100\rangle$ directions. However, this
alters at about 10~K, above which the [110] becomes the most
preferred direction in the system. It is not only that
$M_{\mbox{\tiny {REM}}}$ is the stronger for this direction but
also this remanence is by a factor of $\sqrt{2}$ larger than that
for the $\langle100\rangle$ directions, as is expected from simple
geometric considerations.

\begin{figure}   
\includegraphics[width=7cm]{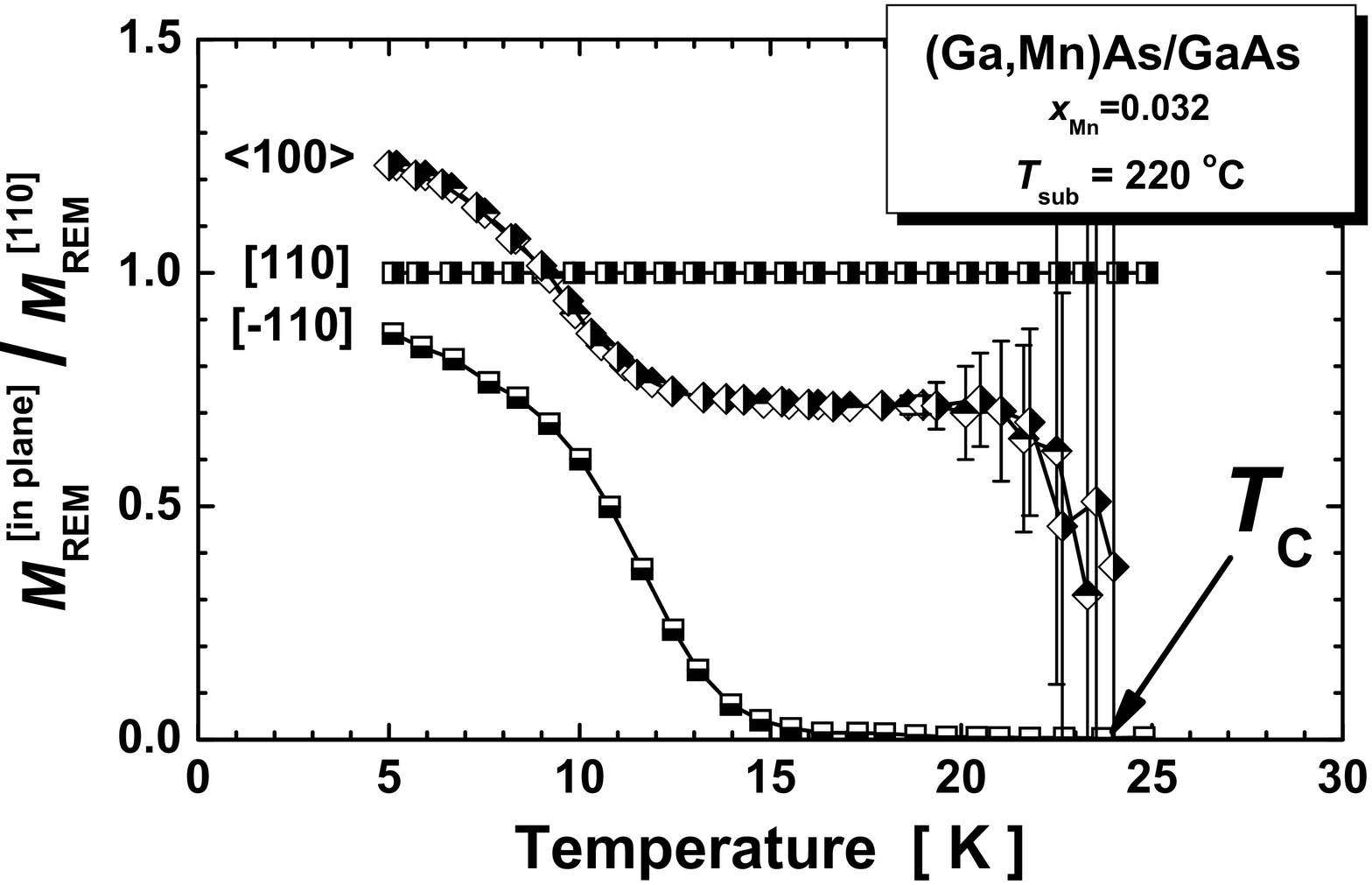}
\caption{\label{fig3}Experimental evidence for the easy axis
switching from $[100]$ to $[110]$ direction (at $\sim10$~K) and
for the uniaxial anisotropy along [110] direction in
Ga$_{0.968}$Mn$_{0.032}$As. The magnetic remanence was measured
for four major in-plane directions and its magnitude is normalized
by the data for the [110] case.}
\end{figure}

Surprisingly, however, the data reveals the existence of an
uniaxial anisotropy along $[110]$. As shown in Fig.~3,
$M_{\mbox{\tiny {REM}}}$ measured along $[\bar{1}10]$ direction
vanishes completely above 15~K indicating that this is the hard
direction of the system. At the same time, the magnitude of
$M_{\mbox{\tiny {REM}}}$ is the same for both $\langle100\rangle$
directions. We argue that this is the result of the lack of the
inversion symmetry in (Ga,Mn)As (001) films. Due to the biaxial
strain, the initial T$_d$ point symmetry of zinc-blende structure
is lowered to D$_{2d}$. But the bottom (Ga,Mn)As/GaAs interface is
different than the top (Ga,Mn)As/vacuum one. This further lowers
the symmetry to C$_{2v}$, where the three principal directions
are: [001], [110] and $[\bar{1}10]$. They are not equivalent, in
C$_{2v}$ the $[110] \Leftrightarrow$ $[\bar{1}10]$\ symmetry gets
broken, while the $[100] \Leftrightarrow [010]$ one is maintained
through reflectivity operations by the $(110)$ and $(\bar{1}10)$
planes, which conforms with the presented results. Since the
cubic-like anisotropy energy is proportional to $M_s^4$ whereas
the uniaxial one to $M_s^2$, the latter is dominating at high
temperatures, where $M_s$ is small. We also note that when
$M^{[\bar{1}10]}_{\mbox{\tiny {REM}}}$ vanishes the
$M^{\langle100\rangle}_{\mbox{\tiny {REM}}}/M^{[110]}_{\mbox{\tiny
{REM}}}$ ratio drops to $\sqrt{2}/2$, the value expected for the
easy axis along $[110]$.

\begin{figure}   
\includegraphics[width=7cm]{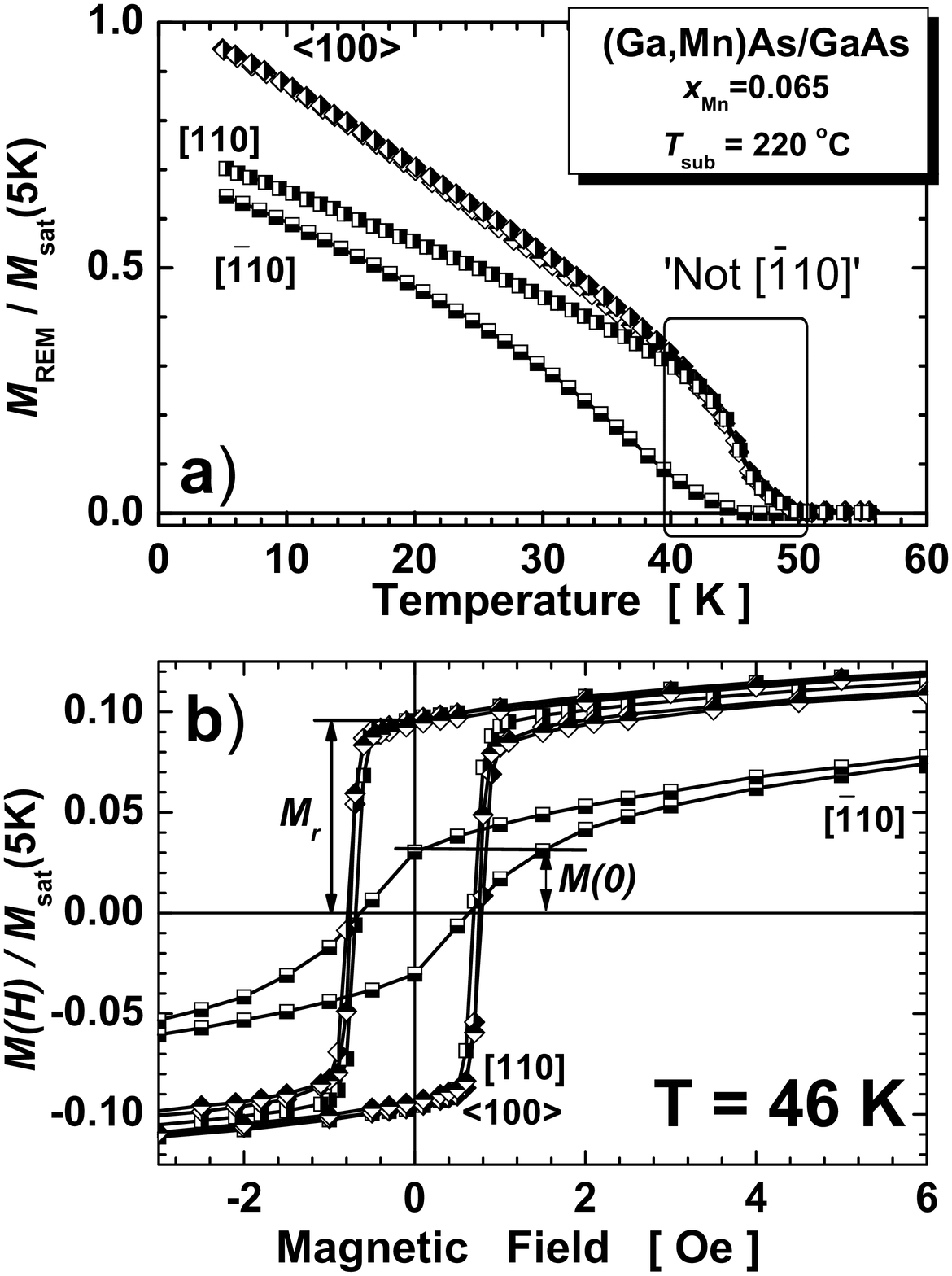}
\caption{\label{Fig4} Magnetization dependencies evidencing "not
$[\bar{1}10]$" anisotropy developing in Ga$_{0.935}$Mn$_{0.065}$As
at temperature close to $T_{\mbox{\tiny C}}$.}
\end{figure}

Another class of samples are those, in which relatively high
values of $p$ and/or $x$ make the cubic easy axis to be oriented
along $\langle100\rangle$ in the whole relevant temperature range.
In such samples, the coexistence of the cubic and uniaxial
anisotropy leads to the case "not $[\bar{1}10]$", as presented in
Fig.~4(a). This is generally the $[100]$ easy axis samples in
which the [110] direction gets equally easy as the $[100]$
directions at high temperatures. So, close to $T_{\mbox{\tiny C}}$
these three major directions are virtually equivalent, while the
$[\bar{1}10]$ axis stays hard. As shown in Fig.~4(b), such an
anisotropy changes considerably the shape of the hysteresis loop,
yielding $M^{[\bar{1}10]}(H=0) = \sqrt{2}M_r / 4$, instead of
zero, as observed in the previous case, for which both cubic and
uniaxial easy axes were along the $[110]$ direction.

In summary, our studies have demonstrated the rich characteristics
of magnetic anisotropies in (Ga,Mn)As/GaAs, which -- in addition
to epitaxial strain -- vary with the hole and Mn concentrations as
well as with the temperature. According to
theory,\cite{Diet01a,Abol01} these reflect spin anisotropy of the
valence band subbands whose shape depends on strain, while their
splitting and population on magnetization and hole concentration.
At the same time, our findings have provided the magnetic
corroboration for the existence of uniaxial in-plane anisotropy.
By group theoretical considerations we have linked this unexpected
anisotropy to the top/bottom symmetry breaking, a suggestion
calling for a microscopic modelling. Furthermore, in view of our
results, a complex temperature dependence of spontaneous
magnetization and of related properties, suggesting --- for
instance --- the presence of phases with different Curie
temperatures, can actually stem from the temperature-induced
switching of the easy axis direction. Since this orientation
depends on the hole concentration which, in turn, can be varied by
the electric field,\cite{Ohno00} it appears possible to switch
magnetization direction in a field-effect transistor with the
channel of a ferromagnetic semiconductor.

\begin{acknowledgments}
The authors thank J. Ferr\'e, H. Ohno and W. Van Roy  for valuable
discussions. Support of the FENIKS project (EC:G5RD-CT-2001-0535)
is gratefully acknowledged.

\end{acknowledgments}



\begin{thebibliography}{99} 

\bibitem{Spintronics}For recent reviews, see, H. Ohno,
F. Matsukura, and Y. Ohno, JSAP International, No. 5, January
2002, pp. 4-13; S. Wolf {\it et al.}, Science {\bf 294}, 1488
(2001); T. Dietl, Acta Phys. Polon. A {\bf 100} (suppl.), 139
(2001).

\bibitem{Mune93} H. Munekata, A. Zaslavsky, P. Fumagalli, and R.J.
Gambino, Appl. Phys. Lett. {\bf 63}, 2929 (1993).

\bibitem{Ohno96Shen97} H. Ohno {\it et al.}, in: {\it Proceedings of the 23rd
International Conference on Physics of Semiconductors}, Berlin
1996, edited by M. Scheffler and R. Zimmermann (World Scientific,
Singapore, 1996) p. 405; A. Shen {\it et al.}, J. Cryst. Growth
{\bf 175/176}, 1069 (1997).

\bibitem{Fedo02}O.M. Fedorych, E.M. Hankiewicz, and Z. Wilamowski,
Phys. Rev. B {\bf 66}, 045201 (2002).

\bibitem{Diet97} T. Dietl, A. Haury, Y. Merle d'Aubign\'e,
 Phys. Rev. {\bf B 55}, R3347 (1997).

\bibitem{Diet00}T. Dietl {\it et al.},  Science {\bf 287}, 1019 (2000).

\bibitem{Diet01a}T. Dietl, H. Ohno, and F. Matsukura,
Phys. Rev. B {\bf 63}, 195205 (2001).

\bibitem{Abol01}M. Abolfath, T. Jungwirth, J. Brum, and A.H.
MacDonald, Phys. Rev. B {\bf 63}, 054418 (2001).

\bibitem{Koni01}J. K\"onig, T. Jungwirth, and A.H. MacDonald,
Phys. Rev. B {\bf 64}, 184423 (2001).

\bibitem{Diet01b}T. Dietl, J. K\"onig, and A.H. MacDonald,
Phys. Rev. B {\bf 64}, 241201(R) (2001).

\bibitem{Kats98}S. Katsumoto {\it et al.}, Phys. Status Solidi (b)
{\bf 205}, 115 (1998).

\bibitem{Gall02}B.L. Gallagher {\it et al.}, (unpublished).

\bibitem{Hrab02}D. Hrabovsky {\it et al.}, Appl. Phys. Lett.
{\bf 81}, 2806 (2002).

\bibitem{Scho01}G.M. Schott, W. Faschinger, and L.W. Molenkamp,
Appl. Phys. Lett. {\bf 79}, 1807 (2001).

\bibitem{Sawi02}M. Sawicki {\it et al.}, \emph{Proc. 2nd Int. Conf. on the Physics and Application of Spin--Related Phenomena in Semiconductors}, Wuerzburg, July 2002, J.
Superconductivity/Novel Magnetism, in press.

\bibitem{Taka02}K. Takamura, F. Matsukura, D. Chiba, and H. Ohno,
Appl. Phys. Lett. {\bf 81}, 2590 (2002).

\bibitem{Endo01}T. Endo, T. S{\l}upi\'nski, S. Yanagi, A. Oiwa, and H.
Munekata, (unpublished).

\bibitem{Ohno00}H. Ohno {\it et al.}, Nature {\bf 408}, 944 (2000).

\end{thebibliography}
\end{document}